\documentclass[12pt]{iopart}

\usepackage{graphicx}
\begin{document}

\title{Evolution of Quantum Criticality  in CeNi$_{9-x}$Cu$_x$Ge$_4$}

\author{L Peyker$^1$, C Gold$^1$, E-W Scheidt$^{1}$, W Scherer$^{1}$, J G Donath$^2$,\\
P Gegenwart$^3$, F Mayr$^4$, V Eyert$^5$, E Bauer$^6$ and H Michor$^6$}

\address{$^1$ CPM, Institut f\"ur Physik, Universit\"at Augsburg, 86135 Augsburg, Germany}
\address{$^2$ Max Planck Institute for Chemical Physics of Solids,
01187 Dresden, Germany}
\address{$^3$ I. Physikalisches Institut, Georg-August-Universit{\"a}t,
37077 G{\"o}ttingen, Germany}
\address{$^4$ EP V, EKM, Institut f\"ur Physik,
Universit\"at Augsburg, 86135 Augsburg, Germany}
\address{$^5$ EP VI, EKM, Institut f\"ur Physik,
Universit\"at Augsburg, 86135 Augsburg, Germany}
\address{$^6$ Institut f\"ur Festk\"orperphysik, Technische Universit{\"a}t Wien,
1040 Wien, Austria}

\eads{\mailto{Ernst-Wilhelm.Scheidt@physik.uni-augsburg.de}, \mailto{Wolfgang.Scherer@physik.uni-augsburg.de}}

\begin{abstract}
Crystal structure, specific heat, thermal expansion, magnetic susceptibility
and electrical resistivity studies of the heavy fermion system
CeNi$_{9-x}$Cu$_x$Ge$_4$ ($0 \leq x \leq 1$) reveal a continuous
tuning of the ground state by Ni/Cu substitution from an
effectively fourfold degenerate non-magnetic Kondo ground
state of CeNi$_9$Ge$_4$ (with pronounced non-Fermi-liquid features)
towards a magnetically ordered, effectively twofold degenerate
ground state in CeNi$_8$CuGe$_4$ with $T_{\rm N } = 175 \pm 5$\,mK.
Quantum critical behavior, $C/T\propto\chi\propto$\,\,-$\ln{T}$,
is observed for $x$\,\,$\cong$\,\,0.4. Hitherto, CeNi$_{9-x}$Cu$_x$Ge$_4$ represents the first
system where a substitution-driven quantum phase transition is
connected not only with changes of the relative strength of
Kondo effect and RKKY interaction, but also with a reduction of the
effective crystal field ground state degeneracy.

\end{abstract}

% \submitto{}
\maketitle

\section{Introduction}
Since the discovery of non-Fermi-liquid (nFL) behavior in
U$_{0.2}$Y$_{0.8}$Pd$_3$ characterized by a logarithmic divergence of the
Sommerfeld coefficient $\gamma\simeq C/T \propto -\ln(T/T_0)$  \cite{Seaman:1991},
the research activity in the field of nFL physics has been very active
\cite{Stewart:2001}. Thereby, great attention was devoted to Kondo systems, in
particular to those where nFL behavior appears to originate from critical
magnetic fluctuations. The latter may emerge near a magnetic phase transition
when a subtle balancing of competing interactions shifts a magnetic phase
transition towards zero kelvin. In this quantum critical point
(QCP) scenario \cite{Hertz:1976,Millis:1993,Moriya:1995},
Kondo interactions, favoring a paramagnetic Fermi-liquid ground state,
compete with RKKY interactions, favoring a magnetically ordered ground
state (for recent reviews, see \cite{Loehneysen:2007, Gegenwart:2008}).

The relative strength of competing Kondo and RKKY interactions, e.g.\ in Ce- or
Yb-intermetallics, can be tuned by parameters such as: (i) pressure
\cite{Bogenberger:1995}, (ii) substitutions \cite{Andraka:1993}, or (iii)
external magnetic fields \cite{Heuser:1998}. Besides these tuning parameters
controlling the relative strength of Kondo and RKKY interactions, there is
another interesting aspect of heavy fermion quantum criticality which was
considered in theoretical studies, but hardly explored experimentally,
namely the variation of the effective degeneracy of total angular momentum
degrees of freedom. This parameter, abbreviated as {\sl effective spin
degeneracy} $N$, which is the number of crystal field (CF) states with energies
up to the magnitude of the Kondo energy, may also drive a system through a QCP.
Coleman \cite{Coleman:1983} has shown that the critical value of the Kondo
coupling constant above which a spin-compensated ground state is stable tends
to zero by $1/N$ as $N$ increases,
i.e.\ systems having a large effective spin degeneracy $N$
are less likely to order magnetically.

In this respect, the heavy fermion CeNi$_{9}$Ge$_4$ represents a suitable model 
system to study the role of effective spin degeneracy since it displays
larger $N$ values than the usual two-fold one in classical nFL systems.
Recently, single ion nFl behavior of the specific heat and magnetic susceptibility
has been discussed for this system. Here, CeNi$_9$Ge$_4$ shows the largest 
ever recorded value of the electronic specific heat coefficient 
$\gamma=C/T \approx$ 5.5\,J\,$\rm K^{-2}mol^{-1}$ at
0.08\,K for paramagnetic Kondo-lattices \cite{Killer:2004, Michor:2004}.
The  dilution of the $f$-moments via Ce/La substitution, i.e.\
Ce$_{1-y}$La$_{y}$Ni$_9$Ge$_4$, revealed an approximate scaling of the magnetic
specific heat contribution and magnetic susceptibility with the
cerium ions  fraction, thus, indicating that the huge
Sommerfeld coefficient $\gamma$ of CeNi$_9$Ge$_4$ is mainly due to Ce single
ion effects, i.e.\ crystal field and Kondo interactions \cite{Killer:2004}.
Another remarkable feature in CeNi$_9$Ge$_4$ is its strongly
temperature-dependent Sommerfeld-Wilson ratio, $R\propto\chi_0/\gamma $, which is revealed
by the distinct different temperature dependencies of specific heat and
magnetic susceptibility below 1\,K \cite{Killer:2004}.

The origin of this behavior is illuminated by CeNi$_9$Ge$_4$ single crystal
susceptibility and polycrystal magnetic entropy data revealing a crystal field scheme of
Ce$^{3+}$ with a quasi-quartet ground state below 20\,K. Thereby, a fourfold effective spin
degeneracy of the Ce-ions is based on two doublets with an energy splitting of
only 0.5\,meV, i.e.\ of  the same order of magnitude as the Kondo energy in
this system, which is about 0.3\,meV \cite{Michor:2006}. Numerical
renormalization group (NRG)-calculations by Anders and Pruschke
\cite{Scheidt:2006, Anders:2006} using the SU(4)-Anderson impurity model which
also accounted for crystal field splitting demonstrated that the Kondo effect
in combination with a quasi-quartet CF ground state leads to a SU(2) to SU(4)
cross-over regime with a significant variation of the Sommerfeld-Wilson ratio as
experimentally observed in Ce$_{1-y}$La$_{y}$Ni$_9$Ge$_4$.

In this work we study the solid solution CeNi$_{9-x}$Cu$_{x}$Ge$_4$, where Ni
is gradually replaced by Cu ions up to $x=1$. This substitution modestly
changes the $3d$-electron number and as a consequence the position of the
Fermi level relative to the Ce-$4f^1$ state. Replacing Ni by Cu is thus
expected to influence the Kondo and RKKY interactions. Usually such
substitution should lower the Kondo temperature and support the formation of
long range magnetic order. The latter effect is also expected as a consequence of an
increasing unit cell volume. Even in the absence of any lattice expansion,
Ni/Cu substitution reduces the local point symmetry at Ce-sites and thus
cancels the quasi-fourfold degeneracy of the CF ground state.

%####################################################################
\begin{figure}
\centerline{\includegraphics[width=9cm,clip]{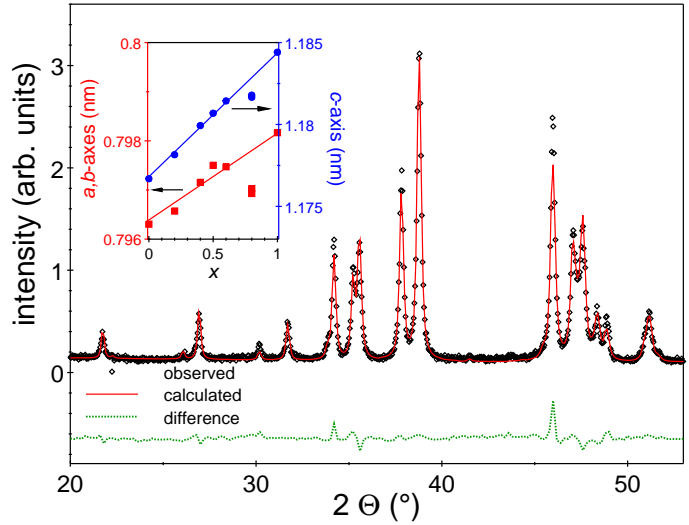}}
\caption{Observed and calculated (Rietveld refinement) x-ray powder diffraction
pattern of annealed CeNi$_8$CuGe$_4$. The doted line displays the difference
plot. The insert shows the variation of the lattice parameters of
CeNi$_{9-x}$Cu$_x$Ge$_4$ with respect to the Cu-concentration \emph{x}.}
\label{fig1}                          % fig-1
\end{figure}
%####################################################################

It is important to mention, that isostructural and isoelectronic CeNi$_9$Si$_4$ with an almost
$4\%$ smaller unit cell volume exhibits distinct Fermi-liquid Kondo lattice
behavior with a about one order of magnitude larger Kondo temperature,
$T_{\rm{K}}\simeq 80$\,K~\cite{michor:2003}, compared to CeNi$_9$Ge$_4$.
X-ray photoelectron spectroscopy on CeNi$_9$Si$_4$ revealed a cerium oxidation state being 
close to $3+$ (in between Ce$^{3.0+}$ and Ce$^{3.1+}$)~\cite{Wang:2007}, 
which suggests that cerium in CeNi$_9$Ge$_4$ is even close to the trivalent state.

\section{\label{sec:cause}Sample preparation and structural characterization}

Polycrystalline samples of CeNi$_{9-x}$Cu$_x$Ge$_4$ and LaNi$_{9-x}$Cu$_x$Ge$_4$ were
prepared by arc-melting of pure elements, Ce: 4N, La: 3N8 (Ames MPC~\cite{ames}),
Ni: 4N5; Cu: 6N; Ge: 5N, under a highly purified argon atmosphere. To obtain the highest possible
homogeneity, the samples were flipped over several times and remelted.
Subsequently, the samples were annealed in evacuated quartz glass tubes for two
weeks at 950$^{\circ}$C. Inductively coupled plasma spectroscopy (ICP-OES) studies were carried out
and confirmed Ni to Cu ratios in good agreement with the relative amount of the
starting materials.

Standard x-ray diffraction techniques using Cu$K_{\alpha}$ radiation were
performed on carefully prepared sieved powdered samples (grain
size 40\,$\mu$m). CeNi$_9$Ge$_4$ crystallizes in the tetragonal space group \emph{I}4/\emph{mcm}
with lattice parameters $a = b = 7.9701(1)\rm{{\AA}}$ and $c = 11.7842(3)\rm{{\AA}}$ (for
structural details, see Ref.~\cite{Killer:2004,Michor:2004}). From Riet\-veld analysis (see as one example the
CeNi$_8$CuGe$_4$ pattern in Fig.~\ref{fig1}) precise lattice parameters of the
solid solutions were determined. The high quality of the refinement ($R_f =
4.14$) is reflected in the difference plot. The analysis indicates that
replacement of the Ni ions by Cu leads to a modest volume expansion, increasing
linearly up to 0.8\,$\%$ for $x =1$ (insert of Fig.~\ref{fig1}). Lattice
parameters of CeNi$_8$CuGe$_4$ are $a = b = 7.9816(7)\rm{{\AA}}$ and $c =
11.8441(2)\rm{{\AA}}$. Total energies calculations based on the new full-potential
augmented spherical wave method \cite{aswbook:2007} suggest some degree of
preferential occupation of the three inequivalent Wyckoff positions 16\emph{k},
16\emph{l}, and 4\emph{d} by Cu: $ E(16k) < E(4d) < E(16l)$. Indeed, the energy
increases for Cu placed on 4\emph{d} and 16\emph{l} sites relative to the
16\emph{k} site are about 0.1\,eV and 0.2\,eV, respectively.

\section{Experimental results}

\subsection{Susceptibility and specific heat}

The temperature dependence of the dc magnetic susceptibility was measured
between 2\,K and 400\,K in an applied magnetic field of 0.5\,T with a commercial
SQUID magnetometer (MPMS7). In the low temperature region ($0.06 \rm{K} < T < 2.5 \rm{K}$)
these measurements were completed by a self designed ac susceptibility device ($B <$\,0.3mT)
installed in a  $^3$He/$^4$He-dilution refrigerator. The absolute values of the low
temperature data were obtained by normalizing the ac-$\chi$ data to the
dc-$\chi$ data between $1.8$\,K and $2.5$\,K. The specific heat experiments were
performed with a commercial equipment (PPMS) between 2\,K and 300\,K
and in a $^3$He-$^4$He-dilution refrigerator down to a base temperature (BT) of 60\,mK
using a standard relaxation method \cite{Bachmann:1972}.

\begin{figure}
\centerline{\includegraphics[width=10cm,clip]{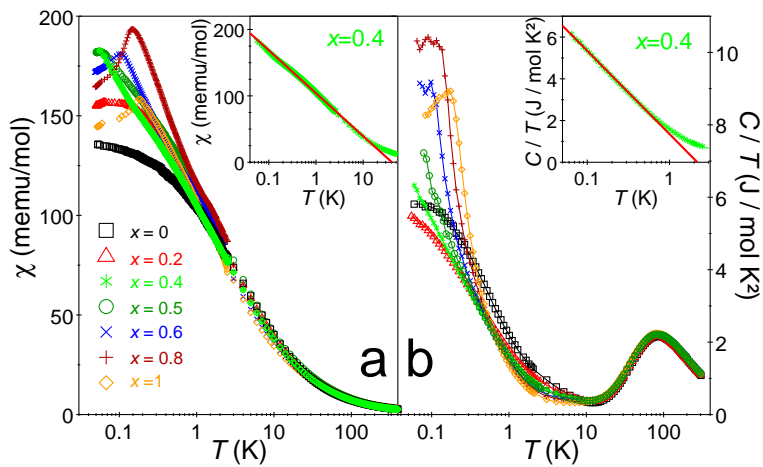}}
\caption{a) The magnetic susceptibility $\chi$ and b) the specific heat divided
by temperature $C/T$ of CeNi$_{9-x}$Cu$_{x}$Ge$_4$ in semi-logarithmic plots.
AFM transitions are evident for $x \geq 0.5$. The inserts show $\chi$ and $C/T$
\emph{vs.} $T$ of CeNi$_{8.6}$Cu$_{0.4}$Ge$_4$: Both solid lines represent
logarithmic fits over more than one decade in temperature.}
\label{fig2}                          % fig-2
\end{figure}

Figure~\ref{fig2} presents the overall susceptibility $\chi(T)$ for
various compositions of CeNi$_{9-x}$Cu$_{x}$Ge$_4$. Above 100\,K all
samples follow a simple modified Curie-Weiss type law, $\chi(T) =
C/(T-\Theta)+\chi_0$, yielding a paramagnetic Curie-Weiss temperature
$\Theta$ around -14\,K, $\chi_0 \cong 0.9$\,memu/mol in reasonable
agreement with the Pauli susceptibility of LaNi$_{9}$Ge$_4$ and a Curie
constant $C$ corresponding to an effective paramagnetic moment of $\cong
2.5 \mu_B$, which is in line with the theoretical value of $2.54 \mu_B$
for a Ce$^{3+}$-ion. Starting from the parent compound CeNi$_{9}$Ge$_4$,
Ni/Cu substitution initially increases the low temperature susceptibility
and reduces the temperature below which it
tends to flatten (from $\simeq$\,1\,K for $x = 0$ to $\simeq$\,0.2\,K
for $x = 0.2$). For CeNi$_{8.6}$Cu$_{0.4}$Ge$_4$ we finally observe a
$\chi(T)\propto -\ln(T)$ behavior down to the BT of 60\,mK
(see insert of Fig.~2a) which is indicative of quantum
criticality. At even higher Cu concentrations sharp cusps in $\chi(T)$
reveal long range antiferromagnetic (AFM) order. Finally,
CeNi$_{8}$CuGe$_4$ exhibits magnetic ordering below $T_{\rm N} \approx
175 \pm 5$\,mK.

%####################################################################
%####################################################################

These observations are corroborated by specific heat results shown as
$C/T$\,\,vs.\,\,$T$ plots for the same compositions of
CeNi$_{9-x}$Cu$_{x}$Ge$_4$ in Fig.~\ref{fig2}b. In comparison to
CeNi$_{9}$Ge$_4$, the initial substitution of Ni by Cu, $x=0.2$, reduces the
$C/T$ values, and the observed deviation from $C/T\propto -\ln(T)$ behavior
(which starts below $250 \pm 10$\,mK for $x=0$) shifts to $150 \pm 10$\,mK.
For CeNi$_{8.6}$Cu$_{0.4}$Ge$_4$, a $C/T\propto-\ln(T)$ divergence of the
Sommerfeld coefficient holds over more than one decade in temperature down to the BT of
60\,mK (see insert of Fig.~2b).
Above $x = 0.4$ magnetic phase transitions
are clearly indicated by specific heat anomalies superimposed on a huge
background due to heavy electrons with Sommerfeld values $C/T$ exceeding
10\,J/mol\,K$^2$ for CeNi$_{8.2}$Cu$_{0.8}$Ge$_4$.

\subsection{Volume thermal expansion}
The volume thermal expansion $\alpha(T)=1/V (\partial V/\partial T)$
is ideally suited to study nFL behavior that results from a QCP,
because $\alpha(T)\propto(\partial S/\partial p)$ directly probes
the pressure dependence of entropy which is accumulated close to the
instability. A theoretical study in terms of a scaling analysis,
suggested that the thermal expansion is far more singular than the
specific heat $C(T)/T$ at any pressure-sensitive QCP
\cite{Zhu:2003}.

We obtained $\alpha(T)$ by means of a high-resolution capacitive
dilatometer (redesigned after Pott and Schefzyik~\cite{Pott:1983})
attached to a $^3$He/$^4$He-dilution refrigerator. Measurements on
selected compositions CeNi$_{9-x}$Cu$_x$Ge$_4$ with $x=0, 0.4$, and
$0.5$ were carried out between $0.08$\,K$<T<4$\,K and in applied
magnetic fields up to 4\,T. The volume thermal expansion $\alpha$ is
given by the sum of the linear thermal expansion coefficients along
three perpendicular directions $a$, $b$, $c$, i.e. $\alpha =
\alpha_a + \alpha_b + \alpha_c$. Assuming isotropic behavior in our
polycrystalline samples, we have obtained the volume expansion as
$\alpha = 3 \times \alpha_a$. For the data in a magnetic field, the
linear expansion coefficient along the direction of the applied
magnetic field has been determined and denoted $\alpha_c$ in the following. 
This estimate does not take into account that texture may play a role in
particular for CeNi$_{9}$Ge$_4$.
%####################################################################
\begin{figure}
  \centerline{\includegraphics[width=10cm,clip]{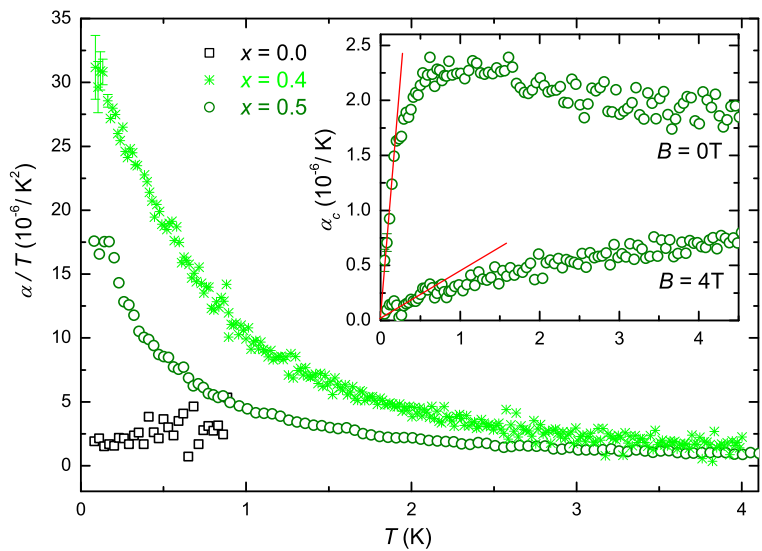}}
  \caption{Volume thermal expansion coefficient
    $\alpha(T)/T$ vs. $T$ of CeNi$_{9-x}$Cu$_{x}$Ge$_4$
    for $x=0, 0.4$, and $0.5$. The errors are indicated 
    by vertical bars.  Insert: Linear thermal expansion
    coefficient $\alpha_c(T)$ of CeNi$_{8.5}$Cu$_{0.5}$Ge$_4$ for $B=0,$
    and $4$\,T. Lines represent fits $\alpha_c(T)=a_1\cdot T$ with
    $a_1=9\cdot 10^{-6}/$\,K$^2$ and $a_1=4\cdot 10^{-7}/$\,K$^2$ for
    $B=0$\,T and $B=4$\,T, respectively.}
\label{fig3}                                   % fig-3
\end{figure}
%####################################################################

Figure~\ref{fig3} shows the volume thermal expansion $\alpha(T)/T$
\emph{vs.} $T$ for various concentrations $x=0, 0.4$, and $0.5$ for
CeNi$_{9-x}$Cu$_x$Ge$_4$. It reaches remarkably high values in the
order of $10^{-6}/$\,K$^2$, typical for heavy fermion compounds. In
agreement with the susceptibility data of the undoped sample, we
find $\alpha(T)/T=$ const for $T<1$\,K indicative for a FL ground
state. Substituting Cu for Ni generates singular behavior in
$\alpha(T)/T$ which is most pronounced for $x=0.4$. Further
increasing the Cu content causes a saturation of $\alpha(T)/T$ at
lowest temperatures $T<0.2$\,K for $x=0.5$. A suppression of the
critical fluctuation is also observed by applying a magnetic field,
which is demonstrated for CeNi$_{8.5}$Cu$_{0.5}$Ge$_4$ in the insert
of Fig. \ref{fig3}. While for $B=0$\,T, FL behavior, i.e. $\alpha_c
(T)/T=$ const is found only up to $T\approx 0.2$\,K, this behavior
extends until $T\approx 0.8$\,K for $B=4$\,T. At the same time, the
linear coefficient $a_1$ of the FL contribution to the thermal
expansion coefficient $\alpha_c(T)\simeq a_1T$ varies by an order of
magnitude from $a_1(B=0$\,T$)=9\cdot 10^{-6}/$\,K$^2$ to
$a_1(B=4$\,T$)=4\cdot 10^{-7}/$\,K$^2$, indicating that $\alpha_c/T$
is strongly suppressed in magnetic field. Such behavior is found in
many quantum critical systems~\cite{Kuchler:2003,Donath:2008}.

\subsection{Electrical resistivity}

\begin{figure}
\centerline{\includegraphics[width=12cm]{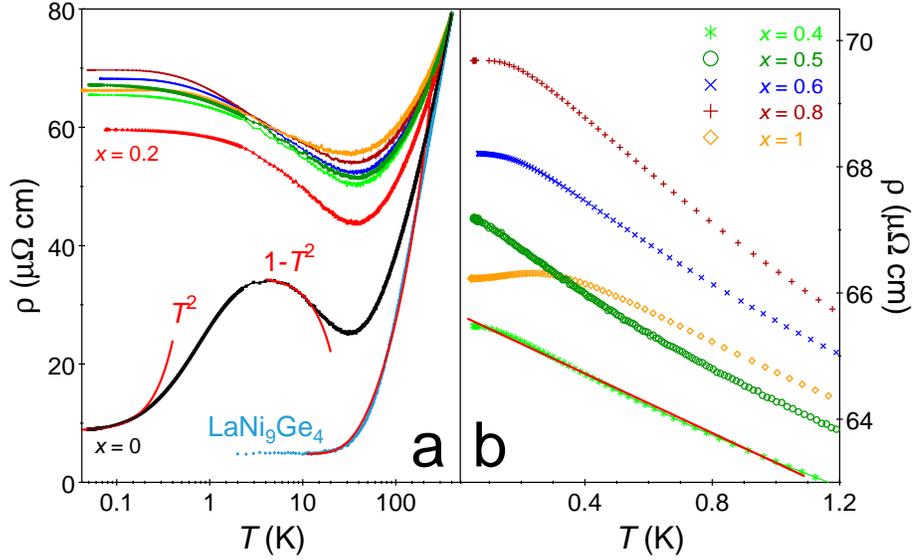}}
\caption{a) A semi-logarithmic plot of the
electrical resistivity $\rho(T)$ of CeNi$_{9-x}$Cu$_{x}$Ge$_4$
normalized at 300\,K to that of LaNi${}_9$Ge${}_4$ which was
measured with the Vander-Pauw method. The solid lines depict
$T^2$-fits for CeNi${}_9$Ge${}_4$ and a Bloch-Gr\"uneisen fit for
LaNi${}_9$Ge${}_4$. CeNi$_9$Ge$_4$ exhibits
Kondo-lattice behavior.  b) In a linear plot of $\rho$ \emph{vs}. $T$
the development of a long range ordered AFM phase transition is
observed  for $x \geq 0.5$, while the resistivity for
CeNi$_{8.6}$Cu$_{0.4}$Ge$_4$ is linear down to 80\,mK (solid line).}
\label{fig4}                          % fig-4
\end{figure}

The concentration dependent crossover from Kondo lattice behavior
with unusual single-ion nFl features of the specific heat and magnetic
susceptibility in CeNi$_9$Ge$_4$ to long range magnetic order in
CeNi$_{9-x}$Cu$_x$Ge$_4$ is also revealed by resistivity
measurements (Fig.~\ref{fig4}a). While the
LaNi$_9$Ge$_4$ reference sample exhibits
a normal metallic Bloch-Gr\"uneisen behavior with a very low
residual resistivity of 5\,$\mu\Omega$\,cm, CeNi$_9$Ge$_4$
seems to represent a classical Kondo lattice, with a residual
resistivity $\rho_0\simeq 9$\,$\mu\Omega$\,cm of a very pure sample.
After passing a minimum around 30\,K and a logarithmic increase,
the resistivity follows a $1-T^2$ law as it is known
for Kondo systems. At lower temperatures the resistivity passes through a
maximum at a temperature $T^{*}\simeq$\,3\,K and follows a $T^2$-behavior
below the Fermi-liquid temperature $T_{\rm{FL}}\simeq$\,160\,mK.  While 
$T_{\rm{FL}}$ is close to the temperature where $C/T$ deviates from the 
$C/T\propto-\ln T$ trend, $T^{*}$ coincides 
approximately with the temperature below which the susceptibility of 
CeNi$_9$Ge$_4$ deviates from the $\chi\propto-\ln T$ behavior
of CeNi$_{8.6}$Cu$_{0.4}$Ge$_4$.

%####################################################################
%####################################################################

For the Cu-substituted samples the resistivity passes through a
Kondo minimum around 30\,K, followed by a logarithmic increase at
lower temperatures. For $x = 0.4$ the resistivity
increases linearly below 1\,K indicating nFL behavior (Fig.~\ref{fig4}b).
This particular behavior was observed for so called disordered Kondo-systems
\cite{Bernal:1995,Miranda:1995}. For $x \geq 0.5$ the resistivity exhibits
a bending at lower T for some samples and a maximum for CeNi$_8$CuGe$_4$
denoting long range magnetic order. This is in line with the evolution
of the AFM transition observed in $C/T$- and $\chi$-measurements.

\section{Discussion}

%####################################################################
\begin{figure}
\centerline{\includegraphics[width=9cm]{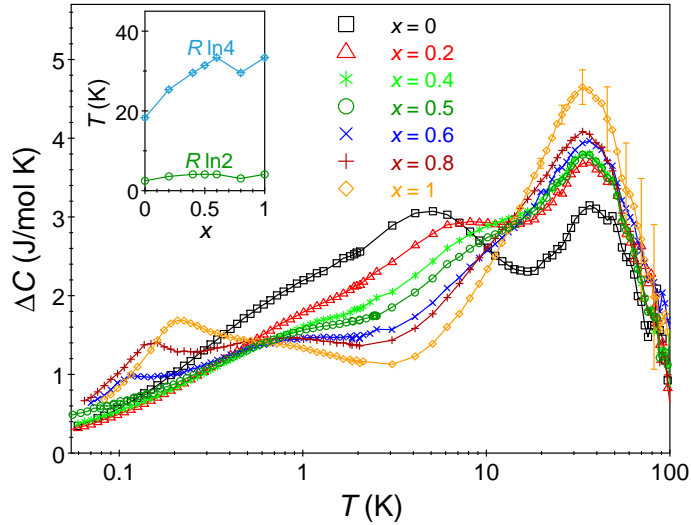}}
\caption{Temperature dependency of the magnetic specific heat $\Delta C$ of
CeNi$_{9-x}$Cu$_{x}$Ge$_4$ in semi-logarithmic representation. The errors 
are indicated by vertical bars exemplary for CeNi$_{8}$CuGe$_4$. The 
resulting temperatures where the entropy 
reaches $R$ln2 and $R$ln4 with respect to the Cu
concentration $x$ are plotted in the insert.}
\label{fig5}                          % fig-5
\end{figure}
%####################################################################

\subsection{Analysis of the high temperature specific heat}

To track the mechanism driving the system from Kondo lattice behavior
with unusual nFl features towards RKKY antiferromagnetism,
we extract the magnetic contribution
of the specific heat by subtracting the total specific heat of the
system LaNi$_{9-x}$Cu$_x$Ge$_4$ with unoccupied $4f$ states.
We therefore synthesized the $4f^0$ reference compounds
LaNi$_{9}$Ge$_4$ and LaNi$_{8}$CuGe$_4$ and
interpolated the total specific heat data of the corresponding
La-sample for each respective composition linearly.

The magnetic contribution to the specific heat $\Delta C$ of all
CeNi$_{9-x}$Cu$_{x}$Ge$_4$ samples is depicted in Fig.~\ref{fig5}. 
The reliability of $\Delta C$ is indicated by vertical error bars  
which become larger at high temperature because of the relatively 
large phonon background. For
CeNi$_{9}$Ge$_4$, two pronounced maxima occur around 5\,K and 35\,K. The former
is associated with the effectively fourfold degenerate Kondo lattice ground
state which is composed by $\Gamma_{7}^{(1)}$ and $\Gamma_{7}^{(2)}$ CF
doublets with an energy splitting of comparable magnitude as the Kondo energy
\cite{Scheidt:2006,Anders:2006}.
Thus, a broad Kondo-like contribution rather than a CF Schottky anomaly becomes
visible. The second, Schottky-like maximum at about 35\,K is associated with a
third CF doublet ($\Gamma_{6}$). With increasing Cu-concentration this
specific heat maximum gains in height, but remains roughly at the same position
near 35\,K. In contrast, the lower Kondo-like maximum decreases and broadens.
At a Cu concentration of $x = 0.4$ a clear separation is observed dividing the
broad hump into a low lying anomaly around 0.8\,K, while the upper anomaly is shifted
towards higher temperature. Finally, for $x=1$, the latter merges with the
Schottky contribution centered at about 35\,K. The appearance of two separated
maxima upon Ni/Cu-substitution, originating from the initially single but broad low
temperature maximum of pure CeNi$_{9}$Ge$_4,$ indicates a reduction of the
effective spin degeneracy of the Ce-ions from fourfold in the case of
CeNi$_{9}$Ge$_4$ to a twofold one for CeNi$_{8.6}$Cu$_{0.4}$Ge$_4$.

The evolution of the temperature dependent magnetic entropy gain, $\Delta S(T),$
further supports a change of energy scales. The insert in Fig.~\ref{fig5} shows
those temperatures where the entropy approaches $R\ln 2$ and $R\ln 4$ in
dependency of the Cu concentration. Both values, $T(S$\,\,=\,\,$R\ln 2)$ and
$T(S$\,\,=\,\,$R\ln 4),$ increase significantly from CeNi$_{9}$Ge$_{4}$ to
CeNi$_{8.4}$Cu$_{0.6}$Ge$_4$, thus, indicating a distinct change of the CF
scheme and/or Kondo energy scale. An increase of the Kondo energy may be
anticipated from the observed increase of $T(S$\,\,=\,\,$R\ln 2)$, but this
trend contradicts the expectation of the change in electron number and unit
cell volume, namely, a reduction of $T_K$ when proceeding from
CeNi$_{9}$Ge$_{4}$ to CeNi$_{8}$CuGe$_4$.

\begin{table}[b]
\caption{\label{tab:table1}
Exchange interaction $J$, the Kondo temperature
$T_{\mathrm{K}}$ as calculated from the  resonant-level model
by Schotte and Schotte~\cite{Schotte:1975} including the
molecular field approximation and the experimentally observed N\'{e}el temperature
$T_{\mathrm{N}}$.}
\begin{indented}
\item[]\begin{tabular}{lccccccccc}
\hline
\hline
 $x$   &&& $J$    &&& $T_{\mathrm{K}}$        &&& $T_{\mathrm{N}}$ \\
       &&&  (K)     &&& (K)                       &&&  (mK)               \\
\hline

     0.5  &&& $<$ 3.9 &&& 2.5 &&&  $55 \pm 10$\\
     0.6  &&& 3.6   &&& 2.2 &&& $100 \pm 5$\\
     0.8  &&& 2.9   &&& 1.7 &&& $135 \pm 5$\\
     1.0  &&& 2.3   &&& 1.3 &&& $175 \pm 5$\\

\hline
\hline
\end{tabular}
\end{indented}
\end{table}

\subsection{Analysis of the low temperature specific heat}

To obtain a more reliable estimate for the trend of the Kondo energies in
CeNi$_{9-x}$Cu$_x$Ge$_4$ we utilized the resonant-level model by
Schotte and Schotte~\cite{Schotte:1975} in combination with a molecular
field approach to account for long-range magnetic order
\cite{Bredl:1978, Gribanov:2006}.

For a spin  $1/2$ system the magnetic contribution of the specific heat $C_{mag}(T)$
follows in this model from:
\begin{equation}
C_{mag} =  2k_{\mathrm{B}}  \mathit{Re} \left\{
\frac{z}{T}
\left[1- \left( \frac{z}{T}-\frac{\partial z}{\partial T} \right)
\psi^{\prime}
\left( \frac{1}{2}-\frac{z}{T} \right)
\right]
\right\},
\label{eq1}
\end{equation}
where $z = T_{\mathrm{K}}+ \mathrm{i} E(T)/2\pi k_{\mathrm{B}}$, with
$T_{\mathrm{K}}$  the Kondo temperature and $E$ the Zeeman energy,
and $\psi^{\prime}$ is the derivative of the digamma function.
By factoring the mean field theory into the resonant-level model, $E$
gets temperature dependent with
\begin{equation}
E(T) = g \mu_{\mathrm{B}} \lambda M(T) = J \frac{M(T)}{g \mu_{\mathrm{B}}}.
\label{eq2}
\end{equation}
Here $g$ is the Land\'{e}-factor (for Ce$^{3+}$\,: $g = 6/7$), $\lambda$ is the
molecular-field constant, $J$ describes the  \emph{s}--\emph{f}
exchange interaction and $M(T)$ is the simple magnetization for a two level system.
Since the levels are broadened by the Kondo effect the simple Brillouin function
becomes modified and $M(T)$ reads:
\begin{equation}
M (T) =  \frac{g \mu_{\mathrm{B}}}{\pi}  \mathit{Im}
\left[
\psi
\left( \frac{1}{2}+ \frac{T_{\mathrm{K}} + \mathrm{i} E(T)}{2 \pi k_{\mathrm{B}} T} \right)
\right]
\label{eq3}
\end{equation}
Finally, while  Eq. (\ref{eq2}) and Eq. (\ref{eq3}) are implicit equations for $E(T)$ and $M(T)$, respectively, we
have to calculate the specific heat (Eq. (\ref{eq1})) numerically.

Model calculations for specific heat data were done for all samples with $x \geq 0.5$.
In this composition range,
$T(S$\,\,=\,\,$R\ln 2)$ and $T(S$\,\,=\,\,$R\ln 4)$ are roughly constant at
values almost twice as large as for CeNi$_9$Ge$_4$ (insert of Fig.~\ref{fig5}).

%####################################################################preliminary
\begin{figure}
\centerline{\includegraphics[width=9cm]{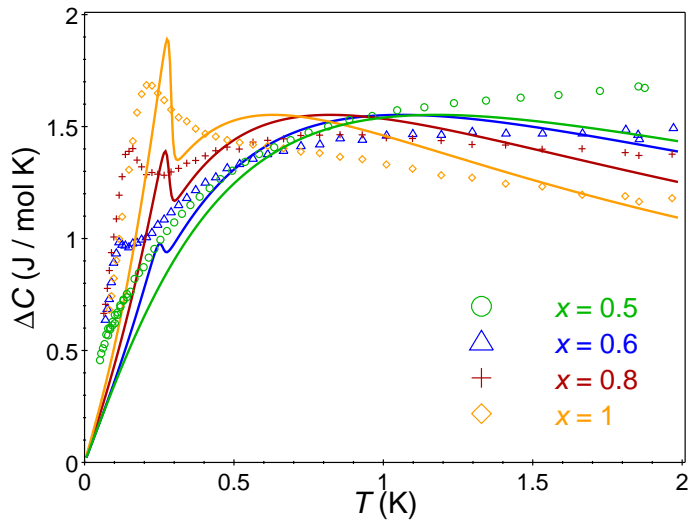}}
\caption{Magnetic contribution to the temperature-dependent specific heat $\Delta C$ of
CeNi$_{9-x}$Cu$_{x}$Ge$_4$ plotted versus $T$ below 2\,K. The solid lines are
fits according to the resonant-level model of Schotte and
Schotte~\cite{Schotte:1975} (see text).}
\label{fig6}                          % fig-6
\end{figure}
%####################################################################

The results are plotted in Fig.~\ref{fig6}. This simple molecular
field model plus the Kondo effect based on a doublet ground state,
of course, does not account for any kind of short range
magnetic correlations or fluctuations. It therefore fails to fit the
experimental data over an extended temperature range. Nevertheless, it
qualitatively reproduces the evolution of the magnetic specific heat anomalies
and Kondo contributions of CeNi$_{9-x}$Cu$_x$Ge$_4$ for $x > 0.5$.
The exchange interactions $J$ and the Kondo temperatures $T_{\mathrm{K}}$
obtained from the resonant-level model are recorded in Tab.~\ref{tab:table1}. In addition,
$T_{\mathrm{K}}$  is depicted in a magnetic phase diagram of CeNi$_{9-x}$Cu$_x$Ge$_4$
(see Sec.~\ref{sec:phase}). A linear extrapolation of the $T_{\mathrm{K}}$
values above $x = 0.5$ suggests a $T_{\mathrm{K}} = 3.5$\,K for CeNi$_{9}$Ge$_4$,
which is in line with $T_{\mathrm{K}}$ revealed by the quasi-elastic line width
observed by cold neutron scattering \cite{Michor:2006}.
The substantial reduction of $T_{\rm{K}}$ with increasing Cu concentration is
in accordance with the usual trend observed in case of Ni/Cu substitution
in other Cerium heavy fermion systems \cite{nieuwenhuys:1995}. 	In addition, the drop of $T_{\rm{K}}$
combined with the lowering of the exchange interaction parameter $J$ is
in accordance with	the Doniach picture \cite{Doniach:1977}.
Therefore, the change of the magnetic entropy gain observed from $x=0$ to
$x=0.5$ is thus attributed to CF effects with a reduction of the effective
spin degeneracy of Ce-ions from fourfold in case of CeNi$_{9}$Ge$_4$ to
a twofold one for CeNi$_{8.6}$Cu$_{0.4}$Ge$_4$.

%####################################################################
\begin{figure}
  \centerline{\includegraphics[width=10cm,clip]{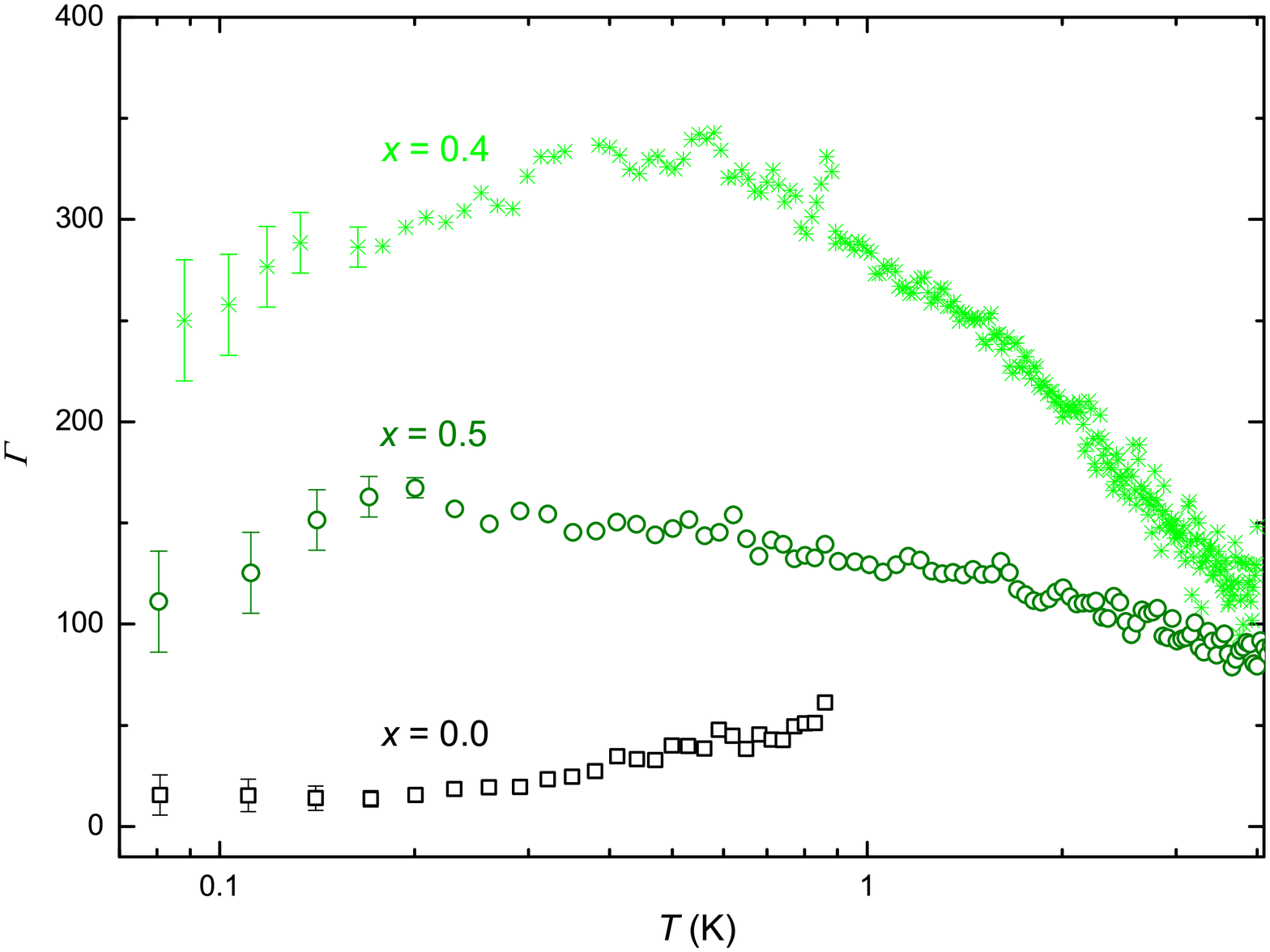}}
  \caption{Temperature dependency of the dimensionless Gr\"uneisen ratio
    $\mathnormal{\Gamma}(T)=V_m/\kappa_T\cdot \alpha(T)/C(T)$ of
CeNi$_{9-x}$Cu$_{x}$Ge$_4$ in semi-logarithmic representation ($V_m=7.485\cdot
    10^{-28}$\,m$^3$, $\kappa_T=1\cdot 10^{-11}$\,Pa$^{-1}$, see text
    for details).}
\label{fig7}                          % fig-7
\end{figure}
%####################################################################

\subsection{Thermal expansion and Gr\"uneisen ratio}

To analyze the nature of the QCP indicated by the thermodynamic data
of CeNi$_{8.6}$Cu$_{0.4}$Ge$_4$ we calculate the dimensionless
Gr\"uneisen ratio $\mathnormal{\Gamma}(T)=(V_m/\kappa_T)\cdot
\alpha(T)/C(T)$ displayed in Fig.~\ref{fig7} for
CeNi$_{9-x}$Cu$_x$Ge$_4$ with $x = 0$, 0.4, and 0.5. In this
calculation of $\mathnormal{\Gamma}(T)$, the molar volume is
$V_m=7.485\cdot 10^{-28}$\,m$^3$ and the isothermal compressibility
is assumed to be $\kappa_T=1\cdot 10^{-11}$\,Pa$^{-1}$ which is a
typical value for heavy fermion systems. The temperature independent
Gr\"uneisen ratio of CeNi$_{9}$Ge$_4$ $[ \mathnormal{\Gamma}(T) = $\,const.\,$]$ below
200\,mK and the enhanced values of $\mathnormal{\Gamma}$ compared to
usual metals characterize CeNi$_{9}$Ge$_4$  as Kondo lattice system
\cite{Takke:1981}. In contrast to the latter system, both
CeNi$_{8.6}$Cu$_{0.4}$Ge$_4$ and CeNi$_{8.5}$Cu$_{0.5}$Ge$_4$
exhibit an order of magnitude higher $\mathnormal{\Gamma}$ values
which are typical for heavy fermion systems close to a magnetic
instability \cite{Kuchler:2003,Kambe:1997,Kuechler:2004}. For the
antiferromagnetic system CeNi$_{8.5}$Cu$_{0.5}$Ge$_4$, a negative
Gr\"uneisen ratio is expected below the Neel-temperature
($T_{\mathrm{N}} = 55$\,mK). The decrease of
$\mathnormal{\Gamma}(T)$ below 0.2\,K may indicate short-range order
above $T_{\mathrm{N}}$. In particular for CeNi$_{8.6}$Cu$_{0.4}$Ge$_4$ the
high $\mathnormal{\Gamma}$ values $[\mathnormal{\Gamma}(0.35 K) =
340]$ and the strong temperature dependence of
$\mathnormal{\Gamma}(T)$ above 0.35~K are quite different to the
parent compound CeNi$_{9}$Ge$_4$ and suggest the vicinity to a QCP.
Below 0.35~K, $\mathnormal{\Gamma}(T)$ saturates and passes a broad
maximum, indicating that quantum critical behavior, i.\,e.\,\,the
divergence of $\mathnormal{\Gamma} (T\rightarrow 0)$
suggested by Zhu et al. \cite{Zhu:2003}, is vanishing at
very low temperatures. This could be explained by assuming that
either the $x=0.4$ system is located somewhat away from the QCP or
that the quantum phase transition is slightly rounded by disorder in
line with the results of the electrical resistivity.
Above 0.35\,K the Gr\"uneisen ratio follows within the experimental
resolution a logarithmic dependence, which clearly deviates from the
scaling prediction for a standard QCP by Zhu {et al.}
\cite{Zhu:2003}. We speculate, that the reduction of the effective
crystal field ground state degeneracy near the quantum phase
transition may modify quantum criticality in our system.

%####################################################################preliminary
\begin{figure}
\centerline{\includegraphics[width=7cm]{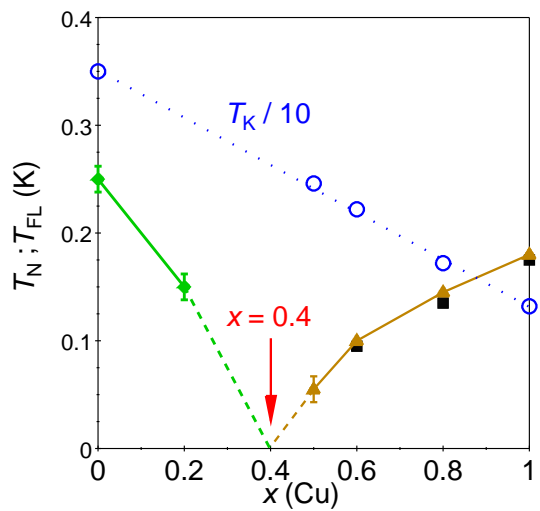}} \caption{
Magnetic phase diagram of CeNi$_{9-x}$Cu$_{x}$Ge$_4$: The squares and triangles
depict $T_{\rm{N}}$ as extracted from the $C/T$ and $\chi$ data, respectively,
while the diamonds represent $T_{\rm{FL}}$ also derived from $C/T$. The Kondo
temperature $T_{\rm{K}}$ (circles) is deduced from the resonant-level model of
Schotte and Schotte~\cite{Schotte:1975} (see Fig.~\ref{fig6} and Tab.~\ref{tab:table1}).}
\label{fig8}                          % fig-8
\end{figure}
%####################################################################

\subsection{Evolution of quantum criticality and crystal field}
\label{sec:phase}

The phase diagram of CeNi$_{9-x}$Cu$_{x}$Ge$_4$ illustrates the presence of
a quantum phase transition near $x = 0.4$ (Fig.~\ref{fig8}). The FL temperature $T_{\rm{FL}}$
is estimated from the deviation from the logarithmic temperature behavior
of the specific heat divided by temperature $C/T$ in the purely
Kondo region. In addition, the N\'{e}el temperature
$T_{\rm{N}}$ is derived from the sharp curvature of the $C/T$
and susceptibility $\chi$ data found in the AFM region. Although the $T_{\rm{FL}}$
values tend to zero around $x \approx 0.4$, the fourfold degenerated
ground state of CeNi$_{9}$Ge$_4$ noticeable splits into two doublets
indicating that samples with $x > 0.4$ exhibit long range
antiferromagnetic order. The $C/T$ and $\chi$ values display logarithmic
temperature dependence over more than one decade in temperature to the BT of 60\,mK at a critical
concentration of $x = 0.4$, while the resistivity $\rho$ displays a
linear $T$-dependence and the thermal expansion coefficient $\alpha / T$
diverges. These results signify the presence of a heavy-fermion QCP.
In CeNi$_{8.6}$Cu$_{0.4}$Ge$_4$, the nFL state develops from a
crossover between a Kondo state ($x \leq 0.4$), where $T_{\rm{FL}}$ tends to zero,
to an antiferromagnetic coherent state ($x \geq 0.4$) starting from
$T_{\rm{N}} = 0$\,K at $x = 0.4$.

%####################################################################
\begin{figure}
\centerline{\includegraphics[width=9cm]{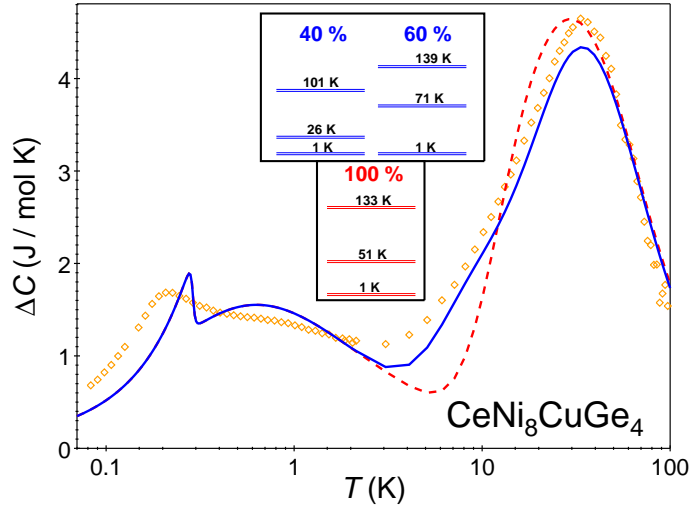}} \caption{
The magnetic contribution of the specific heat $\Delta C$ of CeNi$_{8}$CuGe$_4$
is plotted from 70\,mK to 100\,K. The dashed and solid lines are theoretical
adjustments to the data taking into account the resonant-level model
\cite{Schotte:1975} and two Schottky terms originating from only one (dashed line)
or two (solid line) different CF environments, respectively (see text). }
\label{fig9}                          % fig-9
\end{figure}
%####################################################################

To elucidate the change of the CF scheme in the solid solution from CeNi$_{9}$Ge$_4$
with quasi-fourfold ground state~\cite{Killer:2004,Michor:2006} to CeNi$_{8}$CuGe$_4$
we have quantitatively analyzed the magnetic contribution of the specific heat
$\Delta C$ of CeNi$_{8}$CuGe$_4$ by model calculations which combine specific
CF schemes with an energy split ground state doublet as considered above in the
resonant level model for CeNi$_{8}$CuGe$_4$ with $J=2.3$\,K and $T_{\rm{K}}=1.3$\,K
(see Tab~1).
In Fig.~\ref{fig9} two model cases considering different crystal field
tuning mechanisms are displayed. The first one is based on a
unique crystal field environment for each Cerium atom. Here the CF
splitting of the $J =5/2$ state, $\Delta_{1}=51$\,K is significantly larger
than $\Delta_{1}\sim 6$\,K of the undoped CeNi$_{9}$Ge$_4$, whereas $\Delta_{2} = 133$\,K
is of similar magnitude (see dashed line in Fig.~\ref{fig9}).
The second model scenario considers two or more different CF environments arising from
a stochastic occupation of the Cu atoms on the Ni1 site (Wyckoff position 16\emph{k};
Sec.~\ref{sec:cause}). For example, the solid line in Fig.~\ref{fig9}
represents a model calculation based on two weighted CF-schemes, which
consist of $40\%$ of $\Delta_{1}= 26$\,K, $\Delta_{2} = 101$\,K and
$60\%$ of $\Delta_{1}= 71$\,K, $\Delta_{2} = 139$\,K. This second
model is more consistent with the experimental specific heat
data of CeNi$_{8}$CuGe$_4$ than the model with a unique CF scheme.
In addition, preliminary inelastic neutron scattering (INS) studies indicate
a more complex scheme of CF transitions, thus, supporting the model with more than
one CF environments of Cerium.
Therefore, the nFL behavior of CeNi$_{8.6}$Cu$_{0.4}$Ge$_4$
is attributed to a QCP scenario resulting from both a change
in the CF environment and a non-unique CF scheme. To clarify the
role of chemical disorder upon the CF scheme in these samples,
INS studies are in progress.

\section{Summary}

In conclusion, in the system CeNi$_{9-x}$Cu$_x$Ge$_4$ ($0\leq x \leq
1$) the change from an effectively fourfold degenerate to a twofold
degenerate ground state is accompanied by a quantum phase transition
near CeNi$_{8.6}$Cu$_{0.4}$Ge$_4$ that separates CeNi$_9$Ge$_4$, a
Kondo lattice with unusual nFl features, from the
antiferromagnetically ordered state for $x \geq 0.4$. In this solid
solution Ni/Cu substitution crucially alters the local CF
environment of the Ce-ions. This leads to a quantum phase
transition, which is not only driven by the competition between
Kondo effect and RKKY interaction, but also by a reduction of the
effective crystal field ground state degeneracy.

\section{Acknowledgments}
This work was supported by the Deutsche Forschungsgemeinschaft (DFG)
under Contract No. SCHE487/7-1, the research unit 960 "Quantum phase
transitions" and by the COST P16 ECOM project of the European Union.


\section*{References}
\begin{thebibliography}{10}

\bibitem{Seaman:1991}
Seaman C L, Maple M B, Lee B W, Ghamaty S, Torikachvili M S, Kang J--S, Liu L Z, Allen J W, and Cox L D 1991 {\it Phys. Rev. Lett.} {\bf 67} 2882 (1991).

\bibitem{Stewart:2001}
Stewart G R 2001 {\it Rev. Mod. Phys.} {\bf 73} 797; Stewart G R 2006 {\it Rev. Mod. Phys.} {\bf 78} 743

\bibitem{Hertz:1976}
Hertz J A 1976 {\it Phys. Rev.} B {\bf 14} 1165

\bibitem{Millis:1993}
Millis A J 1993 {\it Phys. Rev.} B {\bf 48} 7183

\bibitem{Moriya:1995}
Moriya T and Takimoto T 1995 {\it J. Phys. Soc. Jpn.} {\bf 64} 960

\bibitem{Loehneysen:2007}
von L\"ohneysen H, Rosch A, Vojta M and W\"olfle P 2007 {\it Rev. Mod. Phys.} {\bf 79} 1015

\bibitem{Gegenwart:2008}
Gegenwart P, Si Q and Steglich F 2008 {\it nature physics} {\bf 4} 186

\bibitem{Bogenberger:1995}
Bogenberger B and von L\"{o}hneysen H 1995 {\it Phys. Rev. Lett.} {\bf 74} 1016

\bibitem{Andraka:1993}
Andraka B and Stewart G R 1993 {\it Phys. Rev.} B  {\bf 47} 3208


\bibitem{Heuser:1998}
Heuser K, Scheidt E--W, Schreiner T and Stewart G R 1998 {\it Phys. Rev.} B {\bf 57} R4198

\bibitem{Coleman:1983}
Coleman P 1983 {\it Phys. Rev.} B {\bf 28} 5255

\bibitem{Killer:2004}
Killer U, Scheidt E--W, Eickerling G, Michor H, Sereni J, Pruschke T and Kehrein S, 2004 {\it Phys. Rev. Lett.} {\bf 93} 216404

\bibitem{Michor:2004}
Michor H, Bauer E, Dusek C, Hilscher G, Rogl P, Chevalier B, Etourneau J, Giester G, Killer U and Scheidt E--W 2004 {\it J. Magn. Magn. Mater.} {\bf 272-276} 227

\bibitem{Michor:2006}
Michor H, Adroja D T, Bauer E, Bewley R, Dobozanov D, Hillier A D, Hilscher G, Killer U, Koza M, Manalo S, Manuel P, Reissner M,  Rogl P, Rotter M and Scheidt E--W 2006 {\it Physica} B {\bf 378-380} 640

\bibitem{Scheidt:2006}
Scheidt E--W, Mayr F, Killer U, Scherer W, Michor H, Bauer E, Kehrein S, Pruschke T and Anders F 2006 {\it Physica} B {\bf 378-380} 154

\bibitem{Anders:2006}
Anders F and Pruschke T 2006 {\it Phys. Rev. Lett.} {\bf 96} 086404

\bibitem{michor:2003}
Michor H, Berger S, El-Hagary M, Paul C, Bauer E, Hilscher G, Rogl P and Giester G 2003
{\it Phys.\,Rev. B} {\bf 67} 224428

\bibitem{Wang:2007}
Wang X, Michor H and Grioni M 2007 {\it Phys. Rev. B} {\bf 75} 035127

\bibitem{ames}
Materials Preparation Center, Ames Laboratory, US DOE Basic Energy Sciences, Ames, IA, USA, available from: $<$www.mpc.ameslab.gov$>$.

\bibitem{aswbook:2007}
Eyert V 2007 {\it The Augmented Spherical Wave Method -- A
Comprehensive Treatment} ({\it Lect. Notes Phys.} vol 719) (Springer, Berlin Heidelberg); Eyert V  {\it J.\ Comp.\ Chem.} in press

\bibitem{Bachmann:1972}
Bachmann R, DiSalvo F J, Geballe T H, Greene R L, Howard R E, King C N, Kirsch H C, Lee K N, Schwall R E, Thomas U-H and Zubeck R B 1972
{\it Rev. Sci. Instrum.} {\bf 43} 205

\bibitem{Zhu:2003}
Zhu L, Garst M, Rosch A, Si Q 2003 {\it Phys. Rev. Lett.} {\bf 91}
066404

\bibitem{Pott:1983}
Pott R and Schefzyik R 1983 {\it J. Phys. Sci. Inst.} {\bf 16} 444

\bibitem{Kuchler:2003}  K\"{u}chler R, Oeschler N, Gegenwart P, Cichorek T, Neumaier K, Tegus O, Geibel C, Mydosh J A, Steglich F, Zhu L, Si
Q 2003 {\it Phys. Rev. Lett.} {\bf 91} 066405

\bibitem{Donath:2008} Donath J G, Steglich F, Bauer E D, Sarrao J L, Gegenwart P 2008
{\it Phys. Rev. Lett.} {\bf 100} 136401


\bibitem{Bernal:1995}
Bernal O, MacLaughlin D E, Lukefahr H G and Andraka B 1995 {\it Phys. Rev. Lett.} {\bf 75} 2023

\bibitem{Miranda:1995}
Miranda E, Dobrosavljevic V and Kotliar G, 1997 {\it Phys. Rev. Lett.} {\bf 78} 290

\bibitem{Schotte:1975}
Schotte K D and Schotte U 1975 {\it Phys. Lett.} A {\bf 55} 38

\bibitem{Bredl:1978}
Bredl C D, Steglich F and Schotte K D 1978 {\it Z. Phys.} B {\bf 29} 327

\bibitem{Gribanov:2006}
Gribanov A, Tursina A, Murashova E, Seropegin Y, Bauer E, Kaldarar H, Lackner R, Michor H, Royanian E, Reissner M and Rogl P 2006  {\it J. Phys.: Condens. Matter} {\bf 18} 9593

\bibitem{nieuwenhuys:1995}
Nieuwenhuys G J 1995 {\sl Handbook of Magnetic Materials} ed. K H J Buschow
(Amsterdam: North-Holland) chapter 1, p 1

\bibitem{Doniach:1977}
Doniach S, 1977 {\it Physica B \& C} {\bf 91} 231

\bibitem{Takke:1981}
Takke R, Niksch M, Assmus W, L\"uthi B, Pott R, Schefzyk R and Wohlleben D K, 1981  {\it J. Phys.: Condens. Matter} {\bf 44} 33

\bibitem{Kambe:1997} Kambe S, Flouquet J, Lejey P, Hean P, de Visser
A 1997 {\it J. Phys.: Condens. Matter} {\bf 9} 4917

\bibitem{Kuechler:2004}
K\"uchler R, Gegenwart P, Heuser K, Scheidt E-W, Stewart G R, Steglich
F 2004 {\it Phys. Rev. Lett.} {\bf 93} 096402



\end{thebibliography}
\end{document}